# Sparse probabilistic evaluation for treatment planning: a feasibility study in IMPT head & neck patients

Jenneke I. de Jong [a][b], Steven J.M. Habraken [c][b], Albin Fredriksson [d], Johan Sundström [d], Erik Engwall [d], Sebastiaan Breedveld [a], Mischa S. Hoogeman [a][b]

  a. Radiotherapy, Erasmus MC Cancer Institute, Rotterdam, The Netherlands
  b. Medical Physics & Informatics, HollandPTC, Delft, The Netherlands
  c. Radiation Oncology, Leiden University Medical Center, Leiden The Netherlands
  d. R&D department, RaySearch laboratories, Stockholm, Sweden

Preprint, March 6, 2026

---

**Abstract**

**Background and purpose**

Probabilistic evaluation improves the trade-off between target coverage and OAR sparing in IMPT but remains computationally demanding. This study proposes sparse probabilistic evaluation (SPE), a computationally efficient approach integrated into a clinical TPS.

**Materials and methods**

Clinical plans of 20 IMPT HNC patients treated in 2024 were included. SPE used a predefined setup and range error grid with Monte Carlo computed dose distributions. Two grid settings were evaluated: the maximum error $E_{max}$ (3σ or 4σ), with $\sigma = \sqrt{(\sigma_{random\,error})^2 + (\sigma_{systematic\,error})^2}$, and the number of setup error points $n_{setup}$ (7, 33, 123).

Accuracy and duration of SPE with each grid were evaluated in the calibration group (5 patients). 1000 treatments with normally distributed random (σ = 1 mm) and systematic (σ = 0.92 mm) setup and range (σ = 1.5%) errors were simulated. The dose distribution of the nearest error point in the grid was assigned to each fraction.

Probability distributions derived from SPE were compared with those from a reference based on 35,000 Monte Carlo calculations. Agreement was quantified using the mean percentile error (MPE), the mean absolute difference across percentiles 0.01, 0.02, …, 1.

The found optimal grid ($E_{max}$ = 3σ, $n_{setup}$ = 33) was applied to the validation group (15 patients).

**Results**

The median MPE in the calibration group decreased significantly as the number of error points increased from 7 ($t_{avg}$ = 2 minutes) to 33 ($t_{avg}$ = 9 minutes), with no further improvement between 33 and 123 ($t_{avg}$ = 27 minutes) error points. Increasing $E_{max}$ from 3σ to 4σ only improved accuracy for values above the 98th percentile.

Applying SPE to the validation group resulted in median errors of 0.02 Gy RBE (range:-0.11 to 0.07) for the 10th percentile of the $D_{99.8\%, CTV}$ distribution and 0.0 Gy RBE (range:-0.14 to 0.23) for the 95th percentile of the $D_{0.03cc, SpinalCord\ Core}$ distribution.

**Conclusion**

Sparse probabilistic evaluation achieves sufficient accuracy while requiring clinically acceptable computation times, paving the way for probabilistic evaluation in clinical practice.

**Introduction**

In proton therapy, mitigating the impact of uncertainties, such as anatomical changes, range errors, patient positioning and beam (setup) errors, is crucial [1]. To account for these uncertainties during treatment plan optimization and evaluation, a common practice is to include the simulation of a set of error scenarios based on systematic setup and range errors [2, 3]. In the Netherlands, scenario-based evaluation has been standardized among the three proton centers, involving the voxel-wise minimum (VWmin) and maximum (VWmax) dose distributions to assess CTV and OAR dose respectively [4]. This approach is based on 28 error scenarios and has been calibrated to maintain population-based consistency in robustness compared to PTV-based photon therapy [4].

However, this method leads to high inter-patient variation in CTV robustness and can be conservative [5]. Probabilistic evaluation [6], which incorporates the probability distributions of the setup and range errors while explicitly modeling both random and systematic errors, could be an improvement. This approach would enable the determination of clinical goal outcomes under specific probabilities. Probabilistic evaluation has been applied for the analysis of the robustness optimization and evaluation methods currently used in clinical practice [5, 7-13]. According to Sterpin et al., explicitly modeling random and systematic errors, coupled with a statistically sound evaluation, improves insight in the actual robustness of the plan for a given confidence level [14]. Furthermore, probabilistic evaluation guided IMPT planning has shown to improve the trade-off between target coverage and OAR sparing in neuro-oncological patients [15].

Nevertheless, the clinical application of probabilistic evaluation is challenging because it typically involves a large number of dose calculations [16, 17]. Recent developments in dose approximation methods have made probabilistic evaluation more computationally feasible. For example, Yang et al used a fast method for probabilistic evaluation that approximated perturbed dose distributions by shifting the nominal dose distribution based on the water equivalent path length [18, 19]. Additionally, Vasquez et al. used deep learning to develop fast percentile-based dose distributions for probabilistic evaluation [20]. Furthermore, fast probabilistic evaluation through polynomial chaos expansion (PCE) [6], based on fitting a polynomial to the dose in every voxel, has been successfully applied in a research setting [5, 7-9, 15].

Despite these advancements, probabilistic evaluation often remains more computationally demanding than the robust evaluation methods currently used in clinical practice [15] and has yet to be integrated into clinical treatment planning systems (TPS).

In this research, we propose a fast method for probabilistic evaluation, which we term sparse probabilistic evaluation (SPE). SPE relies on a predefined setup error grid and range error grid with combined Monte Carlo (MC) computed dose distributions, similarly to PCE. However, instead of fitting a polynomial to the dose in each voxel, SPE uses nearest-neighbor interpolation. This offers a computationally efficient alternative to full probabilistic evaluation, which requires calculating a new dose distribution for every simulated fraction. Additionally, an advantage of SPE is its possible integration with scenario-based optimization. Because scenario-based optimization already computes dose distributions for predefined scenarios, SPE can directly use these distributions to estimate probabilistic outcomes.

In this paper, we investigate whether SPE can accurately reproduce the probability distributions of clinically relevant dose-volume histogram (DVH) parameters. We address this by (l) determining the optimal settings for SPE, balancing computation time and accuracy, and (ll) comparing the outcomes obtained with SPE with a reference for probabilistic evaluation generated by simulating a large number of fractionated treatment courses and computing the individual MC dose distribution for every fraction. This work is conducted in a cohort of head and neck cancer (HNC) patients.

**Materials and methods**

*Clinical treatment plans*

The clinical plans of 20 HNC patients treated at HollandPTC in 2024 were included. The prescribed dose ($D_{pres}$) was 70 Gy RBE for the primary CTV ($CTV_{70.00}$) and 54.25 Gy RBE to the elective lymph nodes ($CTV_{54.25}$), delivered in 2 Gy RBE fractions. The clinical plans were created in RayStation 2023B using the clinical beam model based on the Varian ProBeam system and the MC dose engine with a MC uncertainty of 1% [21].

All clinical plans were generated using IMPT with six beams. Robustness against setup and range errors was aimed for through minimax optimization based on 21 error scenarios [22]. The 21 error scenarios were formed by a combination of a ±3 mm setup error [23] in each of three principal directions and a density uncertainty (range error) of -3%, 0%, or +3%, modelled by scaling the mass density of the patient. The objectives belonging to the min and max CTV dose and the max Mandible dose were set to 'robust', with the aim to achieve the desired outcomes across all error scenarios.

The clinical plans were evaluated based on 28 error scenarios [4, 24]. These error scenarios included 14 setup error combinations with directions determined by the 6 faces and 8 vertices of a cube, all normalized to 3 mm. The setup errors were combined with range errors of -3% or 3%. CTV dose was evaluated using the VWmin dose distribution of the 28 error scenarios, with the clinical goal being VWmin $D_{98\%, CTV}$ = 0.95$D_{pres}$.

*Sparse probabilistic evaluation (SPE)*

SPE relies on a predefined setup error grid and range error grid with Monte Carlo (MC) computed dose distributions. These dose distributions are used to approximate the perturbed dose distributions caused by random and systematic setup and range errors.

We simulated 1000 treatment courses of 35 fractions per patient. For each treatment course, a systematic setup error Σ (mean: 0.00 mm, SD: 0.92 mm) and a range error ρ (mean: 0%, SD: 1.5%) were randomly sampled from their Gaussian distribution and applied across all fractions. Additionally, a randomly sampled random setup error σ (mean: 0.00 mm, SD: 1.00 mm, also Gaussian) was applied to each fraction. The standard deviations (SD) of Σ and σ were chosen to be consistent with the 3.00 mm margin used at our center according to the van Herk recipe (M=2.5·0.92+0.7·1=3) and are close to the SDs found in former measurements performed at our center [9]. The SD of ρ was chosen to be slightly more conservative than the values found in literature [25]. Gaussian distributions are assumed because setup and range errors result from the sum of many small, independent sources.

For each fraction, the combination of Σ and σ was mapped to the nearest error point in the predefined setup error grid. ρ was mapped to the nearest point in the range error grid. The dose distribution corresponding to both the selected setup error point and range error point was then assigned to that fraction.

Two key settings regarding the setup error grid are the maximum error ($E_{max}$) and the grid order. The $E_{max}$ defines the range of possible error points to be considered in the setup error grid, spanning from -$E_{max}$ to +$E_{max}$. The grid order determines how many intermediate error points are sampled within this range. For example, for a standard Cartesian grid with grid order = 5 and $E_{max}$ = 3mm, the 1D error points would be -3 mm, -1.5 mm, 0 mm, 1.5 mm, and 3 mm. Combining these points across all three principal directions results in a cube that contains 5 · 5 · 5 = 125 error points. To avoid including extremely unlikely error points, all points in the cube outside of the sphere with radius $E_{max}$ are omitted, leaving 33 error points in this example. These error points are then combined with seven different range errors, spanning from −4.5% to 4.5% (±3·SD of 1.5%) in 1.5% increments, for the final dose calculations.

SPE was implemented in a research version of RayStation 2024A. The clinical beam model of HollandPTC based on the Varian ProBeam system was used. All calculations were performed on a Dell Precision 7680 workstation with a 13th Gen Intel® Core™ i9-13950HX processor (24 cores, 32 logical processors) and 64 GB of RAM.

*Comparison to a reference for probabilistic evaluation*

The reference for probabilistic evaluation followed the same framework as SPE, simulating 1000 fractionated treatment courses. But instead of mapping to the nearest error point in a grid, the MC dose distributions for all of the 35000 (1000 · 35) fractions were calculated. The 35 fractional doses were summed before extracting 1000 probability distributions of clinically relevant DVH parameters. The probability distributions derived from SPE were compared with those derived from the reference for probabilistic evaluation. Agreement was quantified using the mean percentile error (MPE). For each percentile (0.01, 0.02, …, 1) the absolute difference between the corresponding percentiles of the probability distributions of SPE and the reference for probabilistic evaluation was calculated. The MPE was then defined as the mean of these absolute differences over all percentile levels. It is important to note that the MPE, while useful for analyzing overall agreement, is less clinically relevant when considered on its own. Therefore, errors were specifically evaluated at the 10th, 50th, 95th, 98th and 99.5th percentile to capture the error at clinically relevant points, as identified in discussions with clinicians at our center. Figure 1 illustrates the evaluation algorithm.

*Comparing different settings for SPE in the calibration group*

In the calibration group (5 patients), we compared grid orders of 3, 5, and 7, which corresponded to 7, 33, and 123 included error points ($n_{setup}$), respectively (figure 1.2a). Additionally, we compared $E_{max}$ values of 3σ and 4σ, where $\sigma = \sqrt{(\sigma_{randomerror})^2 + (\sigma_{systematicerror})^2} = \sqrt{1^2 + 0.92^2} = 1.36$ (figure 1.2b). Combining the two settings led to six different setup error grids that were used in SPE and compared to the reference for probabilistic evaluation. The significance of the difference in MPE or error at a specific percentile between the different setup error grids was analyzed using the Wilcoxon signed rank test. For each setup error grid, the computational time required for SPE was recorded for each patient within the SPE script, and the mean execution time over all patients was reported.

In the validation group (15 patients) we applied SPE with the found optimal setup error grid ($E_{max}$ = 3σ, $n_{setup}$ = 33).

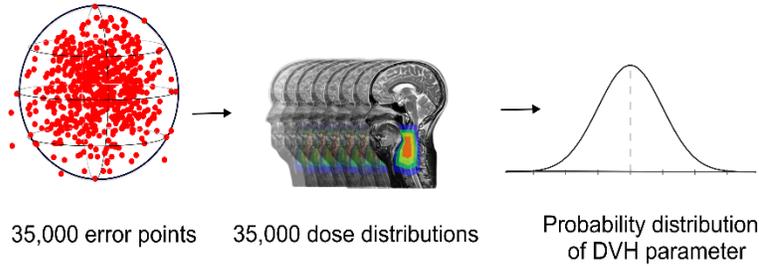
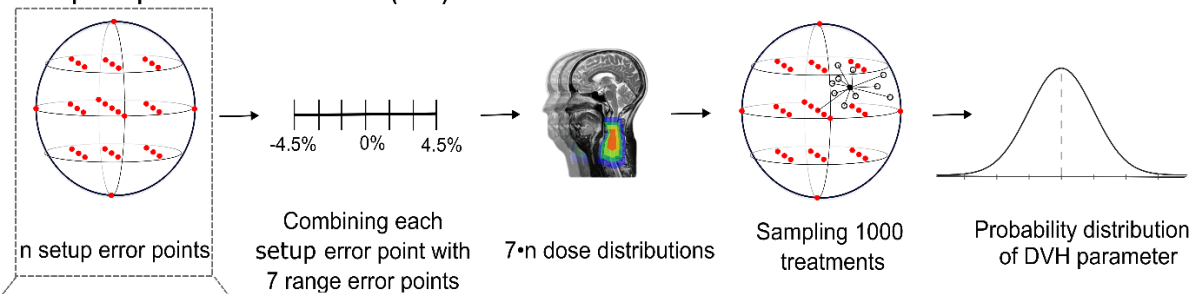
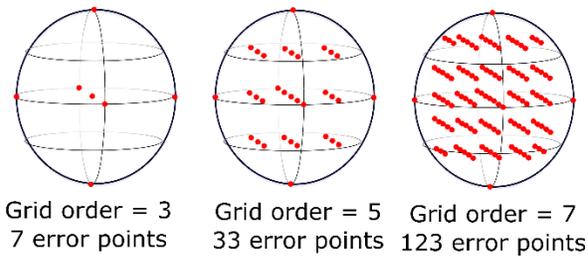
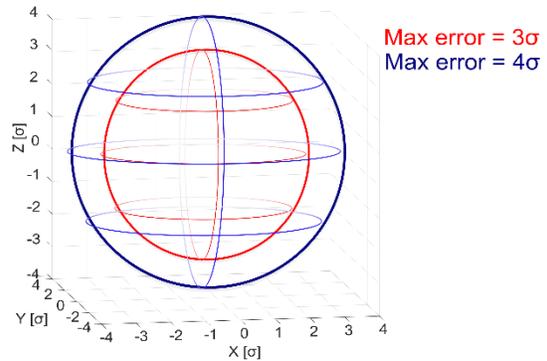

**Fig. 1.** Visualization of the evaluation algorithm for the reference for probabilistic evaluation and sparse probabilistic evaluation. The red dots represent the setup error points. Sparse probabilistic evaluation was tested using setup error grids with 7, 33 and 123 error points ($n_{setup}$) (2a) and maximum errors ($E_{max}$) of $3\sigma$ and $4\sigma$ (2b).

## Results

Figures 2 and 3 present the results for the patients in the calibration group. Figure 2 illustrates the trade-off between accuracy and the number of error points in the setup error grid. The population median MPE decreased for all DVH parameters as $n_{setup}$ increased from 7 to 33. Significant reductions in MPE (p = 0.03) were observed for all DVH parameters, except for the $D_{99.8\%, CTV}$ ($E_{max}$ = 3σ, 4σ), $D_{mean, oral\ cavity}$ ($E_{max}$ = 3σ), $D_{mean, larynx\ SG}$ ($E_{max}$ = 3σ) and $D_{mean, musc\ constrict\ I}$ ($E_{max}$ = 4σ). Increasing the number of error points from 33 to 123 did not yield further improvements for most DVH parameters. Significant reductions in MPE were observed only for $D_{99.8\%, CTV}$ (p = 0.03, $E_{max}$ = 4σ) and $D_{mean, larynx\ SG}$ (p = 0.03, $E_{max}$ = 4σ). Increasing $E_{max}$ from 3σ to 4σ did not significantly reduce the MPE for any DVH parameters at any $n_{setup}$.

The population median computation times of the MC dose distributions and the probability distributions of the DVH parameters for 7, 33 and 123 error points were 4, 9, and 29 minutes, respectively. In supplementary materials 4, the full boxplots of the computation times are shown. In supplementary materials 2, the inter-reference variation, derived by creating the reference for probabilistic evaluation twice for the same patient, is plotted alongside the boxplots in figure 7.

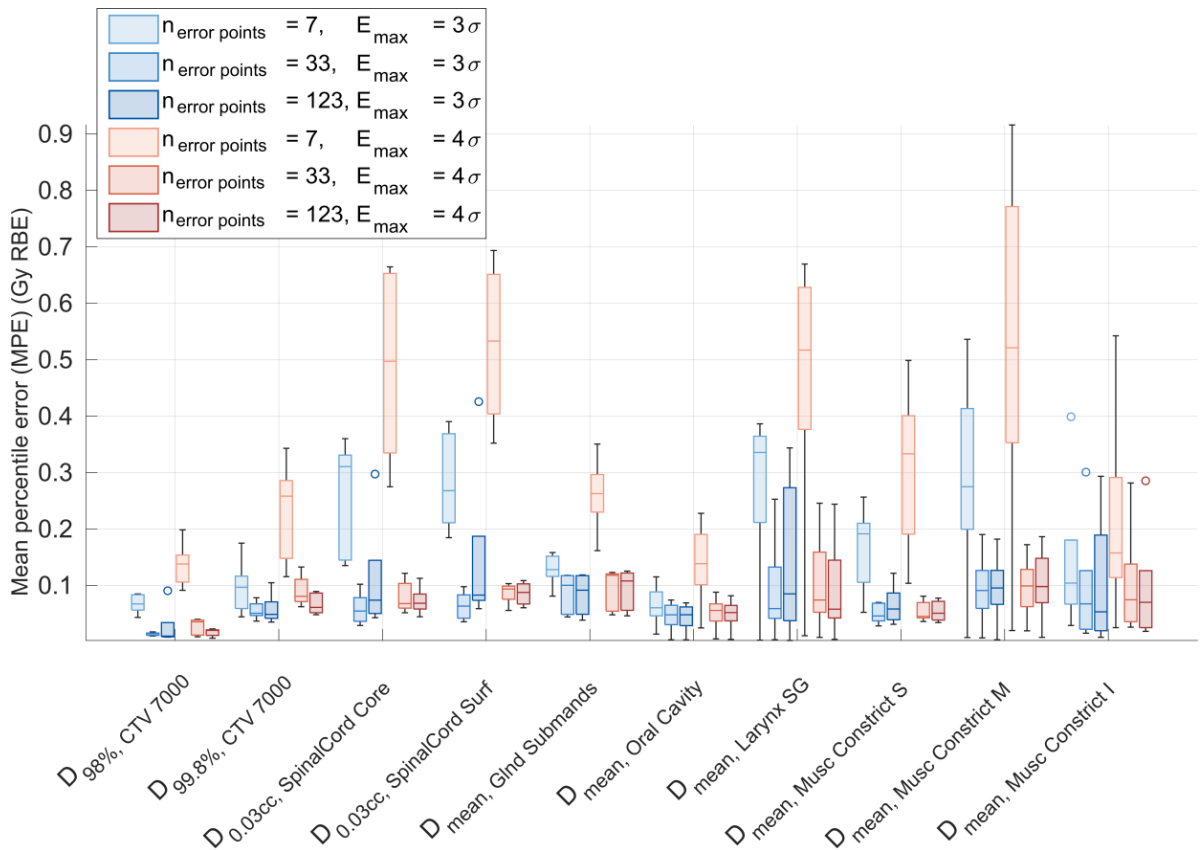

**Fig. 2.** The influence of the included error points ($n_{setup}$) and the maximum error ($E_{max}$) in the setup error grid on the mean percentile error (MPE) in the calibration group (5 patients).

Figure 3 presents the errors for $n_{setup}$ = 33 when comparing $E_{max}$ values of 3σ and 4σ at several extreme percentiles. For the 95th percentile errors, increasing $E_{max}$ to 4σ does not lead to an improvement. For the 98th percentile errors, a significant improvement is observed only for the $D_{mean, musc\ constrict\ m}$ (p = 0.03). For the 99.5th percentile error, median improvements are seen for all OARs; however, these improvements reach statistical significance only for $D_{mean, glnd\ submands}$ (p = 0.03) and the $D_{mean, musc\ constrict\ s}$ (p = 0.03).

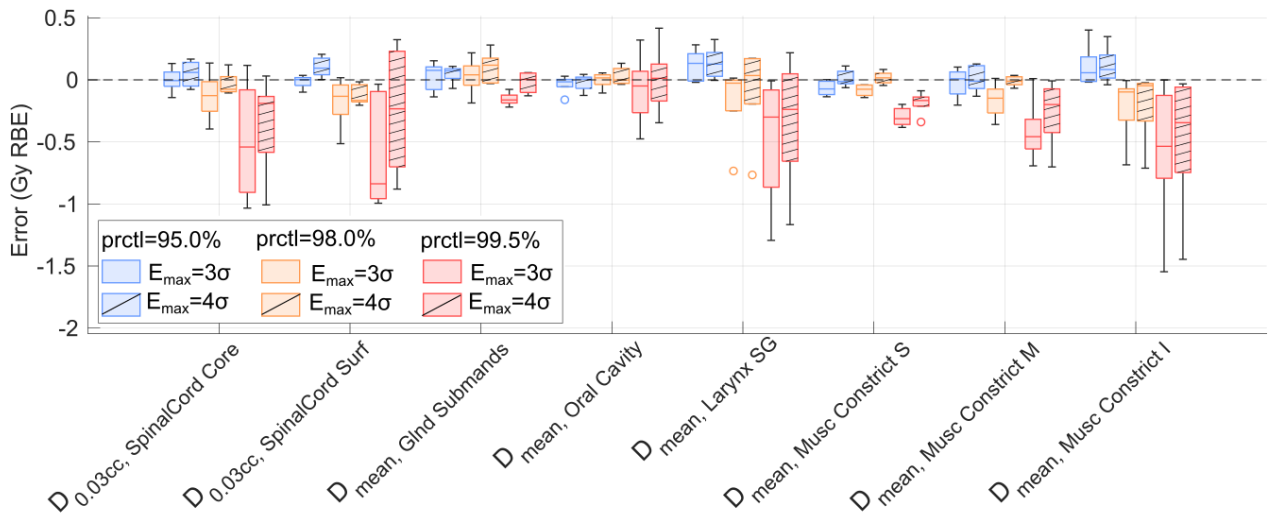

**Fig. 3.** The influence of the maximum error ($E_{max}$) in the setup error grid on errors at selected percentiles of the probability distributions in the calibration group (5 patients).

Figure 4 presents the full probability distributions for a representative patient. Some discrepancies between SPE and the reference for probabilistic evaluation are noticeable. In the probability distributions belonging to $n_{setup}$ = 7, exaggerated peaks around the values belonging to the nominal dose distribution are observed in the probability distributions of the $D_{99.8\%, CTV}$ and the $D_{0.03cc, spinalcord\ core}$. Additionally, in the probability distributions belonging to $E_{max}$ = 3σ, an underestimation of the highest percentile values can be seen in the probability distribution of the $D_{0.03cc, spinalcord\ core}$.

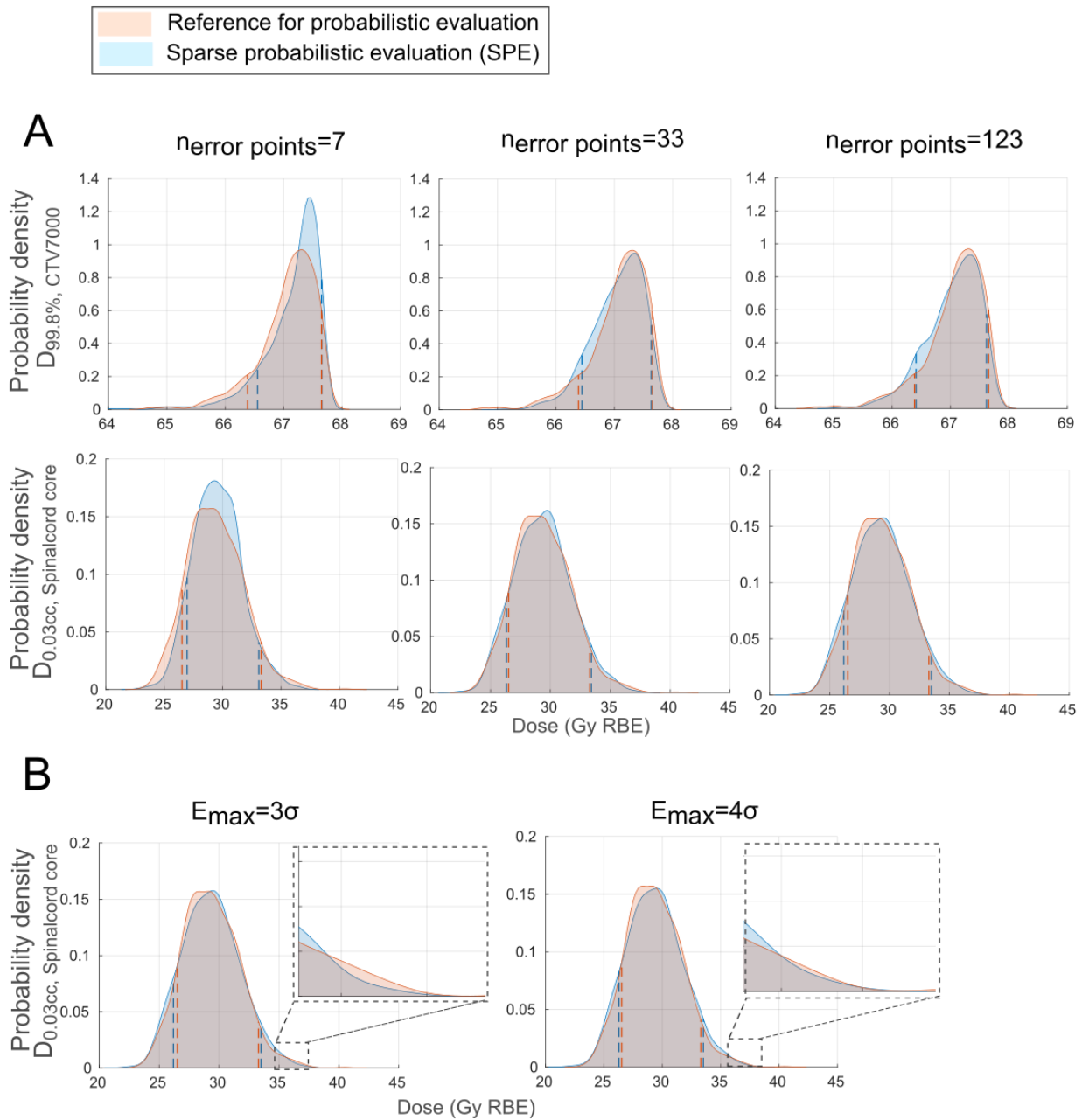

**Fig. 4.** A. The influence of the included error points ($n_{setup}$) in the setup error grid on the probability distributions of the DVH metrics related to robust near-min dose in the CTV and robust near-max dose to the spinal cord in a representative patient. The maximum error ($E_{max}$) is set to 3σ. B. The influence of $E_{max}$ on the probability distributions of the DVH metrics related to robust near-max dose to the spinal cord. The right tail of the distribution is magnified because this is where most influence of the $E_{max}$ setting is expected. In these plots $n_{setup}$ is set to 33.

Figure 5 shows the results for the patients in the validation group for SPE with $E_{max} = 3\sigma$ and $n_{setup} = 33$. The patient population median MPE ranges from 0.01 to 0.08 Gy RBE (supplementary materials 1). The 50th percentile errors are small and centered around zero. For the 95th percentile values we see a small positive shift for the $D_{mean, Larynx\ SG}$ (population median: 0.04 Gy RBE) and the $D_{mean, Musc\ constrict\ I}$ (population median: 0.11 Gy RBE). In contrast, for the 98th percentile values we see a small negative shift for almost all DVH parameters (population median: -0.15 to 0.00 Gy RBE).

Focusing on two clinically relevant points, SPE results in median errors of 0.02 Gy RBE (range:-0.11 to 0.07 Gy RBE) for the 10th percentile of the $D_{99.8\%,\ CTV}$ distribution and 0.0 Gy RBE (range:-0.14 to 0.23 Gy RBE) for the 95th percentile of the $D_{0.03cc,\ SpinalCord\ Core}$ distribution.

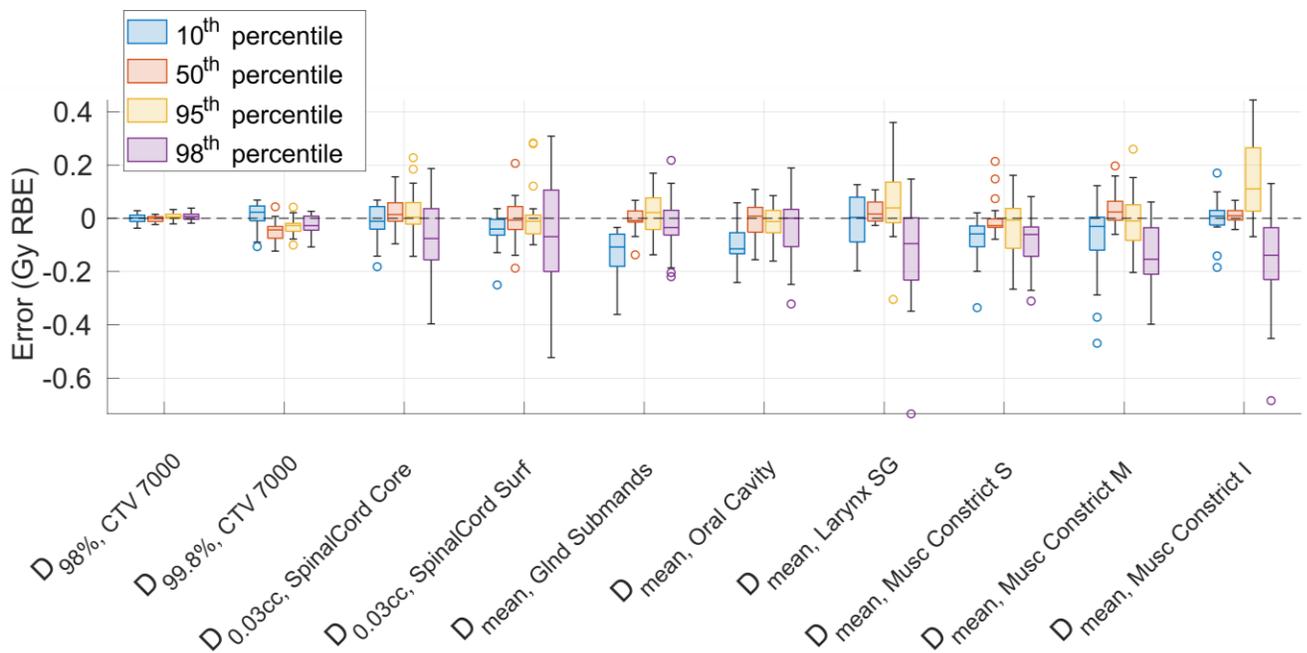

**Fig. 5.** The error in the validation group (15 patients) for clinically relevant percentiles when applying SPE with 33 included error points ($n_{setup}$) and a maximum error ($E_{max}$) of $3\sigma$.

## Discussion

In this paper, we determined the optimal settings for SPE by comparing its results to a reference for probabilistic evaluation generated by large-scale MC dose calculations. Our main finding is that SPE performed in 9 minutes leads to accurate estimations of probabilistic DVH values, with under- and overestimations within clinical acceptability, supporting the feasibility of applying SPE in clinical practice.

Analysis of SPE with different settings led to three key findings. First, increasing the number of error points in the setup error grid from 7 to 33 resulted in significant improvement for the majority of DVH parameters, but increasing further to 123 errors points did not. This suggests that 33 points are sufficient to capture the general shapes of the probability distributions. However, when only systematic errors are included, the distributions do exhibit spikiness. Smoothing of the SPE probability distribution occurs when random errors are sampled for each treatment course and averaged, which aligns the distributions with the reference for probabilistic evaluation.

Second, the errors plateaued at values greater than zero. This is partly caused by the limited precision of the reference for probabilistic evaluation, which was based on 1,000 treatment course simulations. While increasing the number of simulations to 10,000 could improve accuracy, this was computationally infeasible. Repeated creation of the reference for probabilistic evaluation with a different seed for the random sampling of setup and range errors confirmed this, as the inter-reference variation aligned with the observed plateau (Supplementary Materials 2). Other causes could be the inclusion of only seven range error points and the MC noise.

Third, extending $E_{max} = 3\sigma$ to $E_{max} = 4\sigma$ resulted in improvements only for the extreme percentiles (e.g. 99.5). This is expected, as the tails of the probability distributions are predominantly influenced by the most extreme error points. For $n_{setup} = 7$, extending $E_{max} = 3\sigma$ to $E_{max} = 4\sigma$ led to a worse MPE because the sampling of the extreme error points reduced the sampling density near zero, where the probability mass is greatest.

We applied SPE with $E_{max} = 3\sigma$ and $n_{setup} = 33$ to the validation group. Errors at the 50th percentile were generally small and centered around zero, reflecting the low sensitivity of mid-range percentiles. In contrast, the 98th percentile values were often slightly underestimated (population median: -0.15 to 0.00 Gy RBE), as the exclusion of errors beyond $3\sigma$ fails to account for large deviations that significantly influence extreme percentiles. Lastly, the 95th percentile values were often slightly overestimated for the $D_{mean, Larynx\ SG}$ (population median: 0.04 Gy RBE) and the $D_{mean, Musc\ constrict\ I}$ (population median: 0.11 Gy RBE). This overestimation can be attributed to spikiness in the sparse evaluation probability distributions (Supplementary materials 3), as these structures are highly sensitive to range errors and the inclusion of only seven range error values was insufficient to completely smooth the probability distributions.

For this study, we limited ourselves to analyzing the settings of the setup error grid, combining each grid with seven range error points. An improvement could involve a combined setup and range error grid and the modeling of the combined probability of both setup and range errors, enabling the filtering of combined error points based on a 4D probability threshold. This approach may be able to reduce the number of combined error points without significant information loss and allow for the inclusion of more than seven range error points for certain setup error points without increasing computational time. Additionally, we assumed independent Gaussian-distributed setup errors along the x, y, and z axes. However, in clinical practice, these errors may exhibit correlations, as body shifts or rotations can simultaneously affect multiple axes. Lastly, we used a simple Cartesian grid for error point selection, but alternative grid structures may provide more efficient coverage of the error space.

Future research should explore probabilistic optimization methods [26-28] to directly steer towards probabilistic outcomes during optimization. These methods could incorporate error grids and sampling strategies similar to those used in this study.

Before SPE can be implemented in clinical practice, it should be evaluated for other treatment sites, as accuracy likely depends on several treatment site characteristics. For sites with fewer fractions or where random errors are small relative to systematic errors, the probability distributions may exhibit greater variability and spikiness. Accuracy is also influenced by structure and voxel size, with larger structures such as the CTV showing higher accuracy than smaller structures like the larynx. The magnitude of the standard deviations of setup and range errors are also expected to influence accuracy, with larger errors leading to reduced accuracy.

Collaboration with clinicians and medical physics experts will be essential to determine which percentiles of the probability distributions are most relevant for decision-making and to establish acceptable accuracy thresholds for each treatment site.

In conclusion, SPE is a probabilistic evaluation method that achieves high accuracy while not requiring a substantially larger computation time compared to current robustness evaluation methods. This paves the way for the application of probabilistic evaluation in clinical practice.

**CRediT authorship contribution statement**

Jenneke I. de Jong**:** Conceptualization, Methodology, Validation, Formal analysis, Investigation, Writing – original draft, Writing – review & editing, Visualization.

Steven J.M. Habraken:  Conceptualization, Methodology, Validation, Resources, Writing – review & editing, Supervision.

Sebastiaan Breedveld: Conceptualization, Methodology, Validation, Writing – review & editing, Supervision.

Albin Fredriksson: Conceptualization, Methodology, Software, Writing – review & editing.

Johan Sundström: Conceptualization, Methodology, Software, Writing – review & editing.

Erik Engwall: Conceptualization, Methodology, Software, Writing – review & editing.

 Mischa S. Hoogeman: Conceptualization, Resources, Writing – review & editing, Supervision, Project administration, Funding acquisition.

**Funding**

The research for this work was partly funded by RaySearch Laboratories and partly funded by the Surcharge for Top Consortia for Knowledge and Innovation (TKIs) from the Dutch Ministry of Economic Affairs and Climate.

The Rene Vogels Grant facilitated a three-month visit to collaborate with the optimization team of RaySearch Laboratories in Stockholm.

**Declaration of competing interest**

The authors declare that they have no known competing financial interests or personal relationships that could have appeared to influence the work reported in this paper.

**Supplementary materials 1**

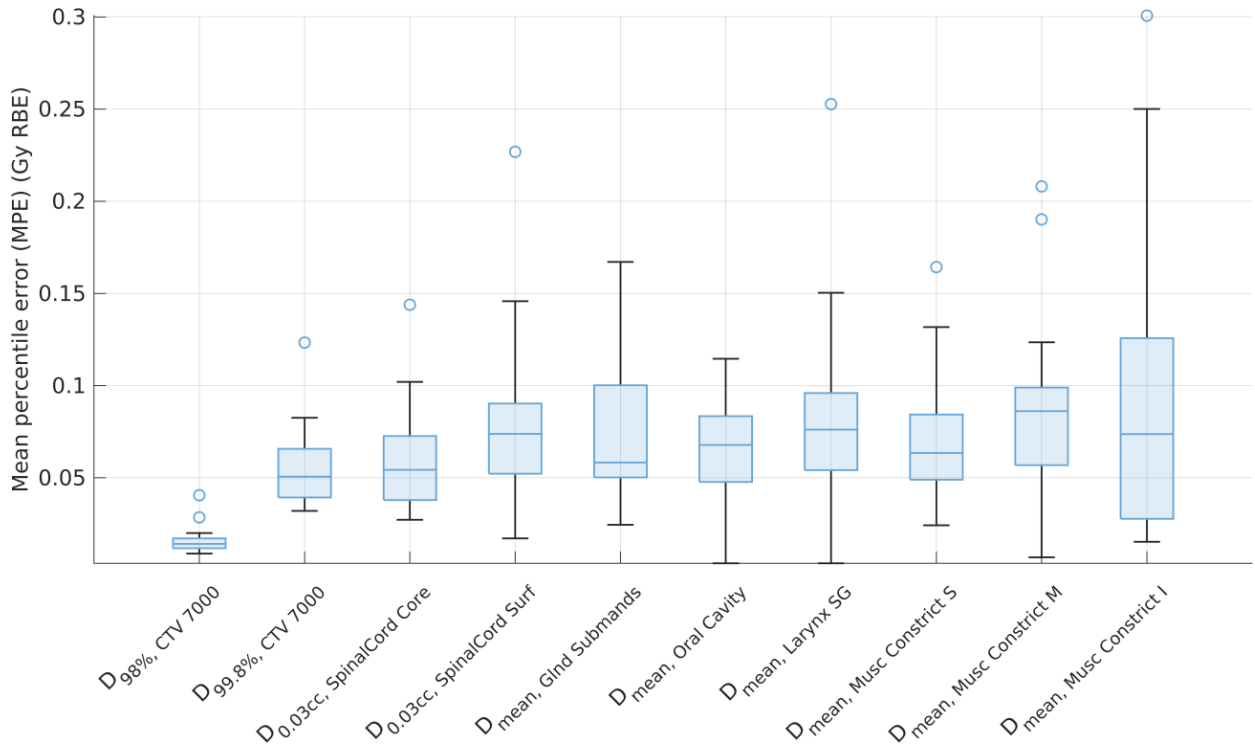

**Fig. 6.** The mean percentile error (MPE) in the validation group (15 patients) when applying SPE with 33 included error points ($n_{setup}$) and a maximum error ($E_{max}$) of $3\sigma$.

**Supplementary materials 2**

Figure 7 and 8 present the inter-reference variation resulting from repeating the generation of the reference for probabilistic evaluation based on 35,000 MC–calculated dose distributions using a different seed for the random sampling of setup and range errors. The Monte Carlo noise seed was kept constant.

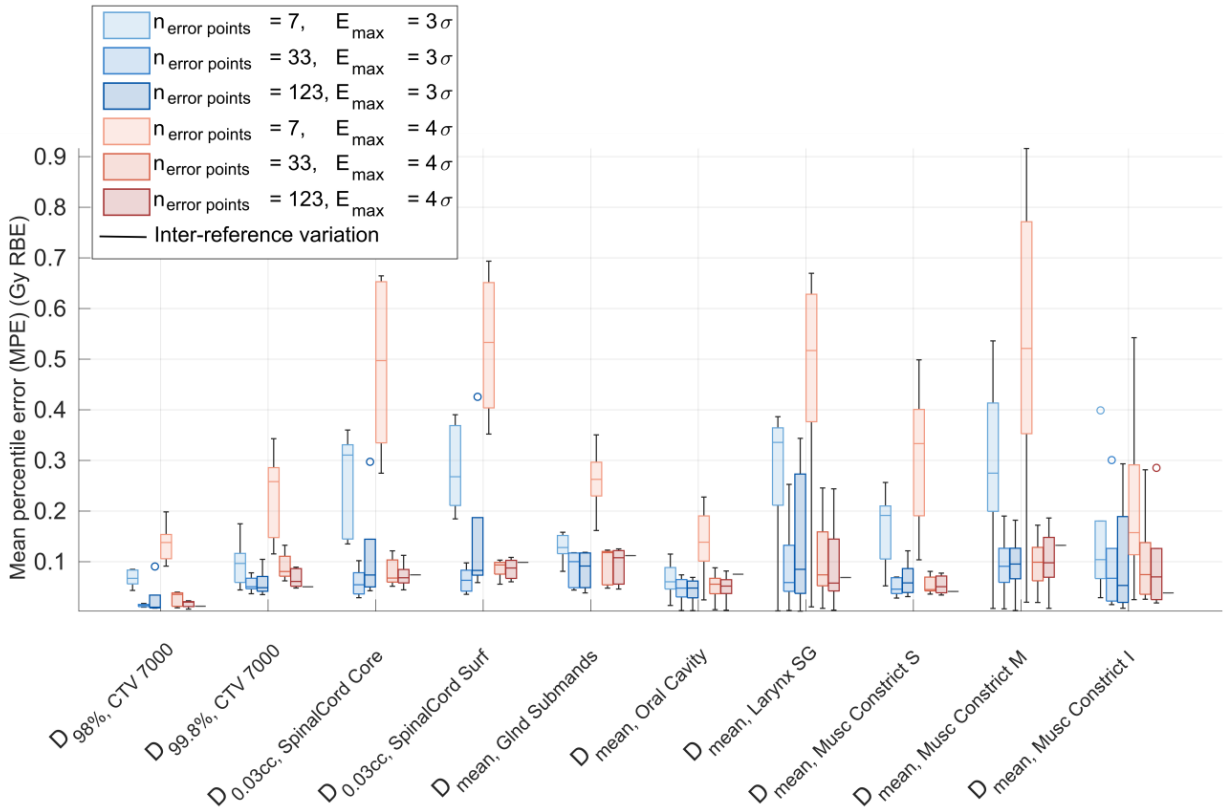

**Fig. 7.** The influence of the number of error points ($n_{setup}$) and the maximum error ($E_{max}$) in the setup error grid on the mean percentile error (MPE) in the calibration group (5 patients). The inter-reference variation was assessed by creating the reference for probabilistic evaluation, based on 35,000 MC–calculated dose distributions, twice and is plotted next to the boxplots using black horizontal lines.

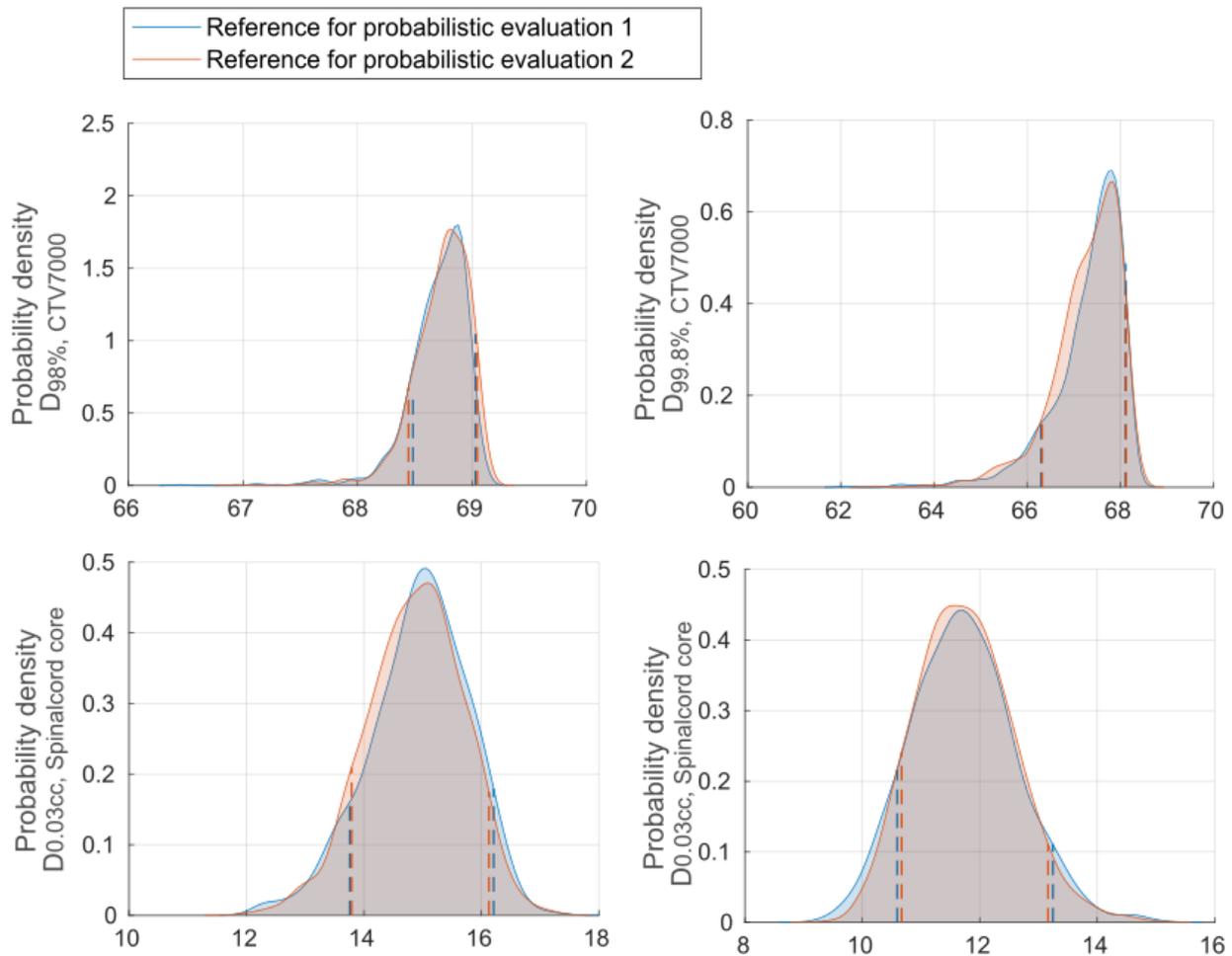

**Fig. 8.** The inter-reference variation when repeating the method for generating the reference for probabilistic evaluation based on 35,000 MC–calculated dose distributions in the probability distributions of the DVH metrics related to robust near-min dose in the CTV and robust near-max dose to the spinal cord in a representative patient. The resulting probability distributions do not match perfectly, indicating that the reference estimation retains some degree of stochastic variability.

**Supplementary materials 3**

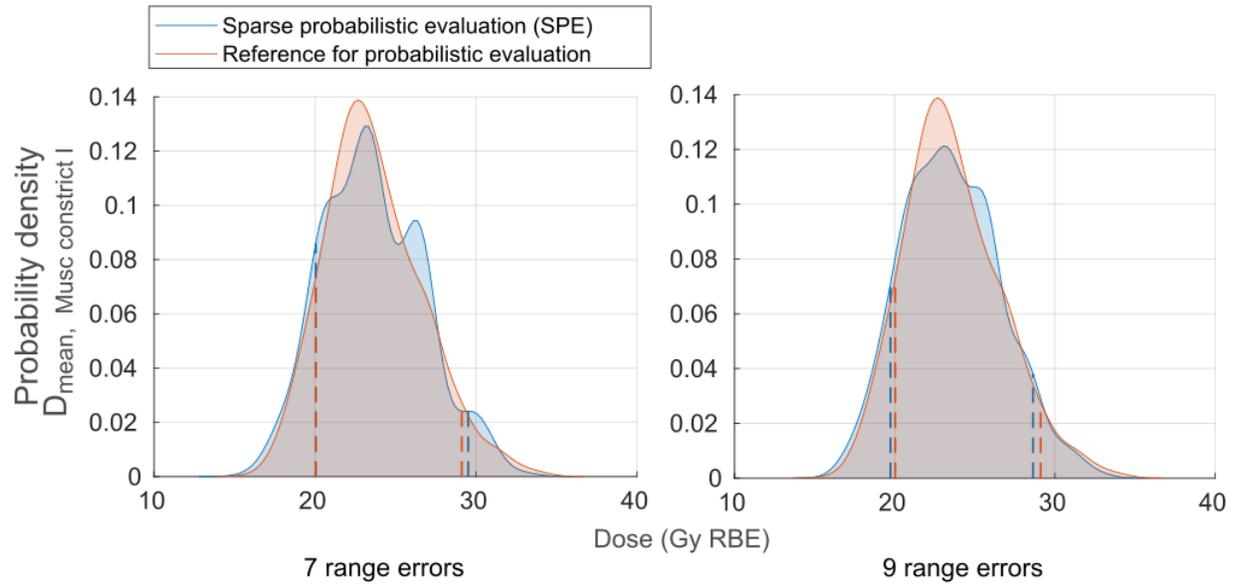

**Fig. 9.** Illustration of the results observed in some patients for structures highly sensitive to range uncertainties when applying SPE with 33 included error points ($n_{setup}$) and a maximum error ($E_{max}$) of $3\sigma$. The figure shows the probability distribution of the $D_{mean}$ of the interior constrictor muscle with the right dotted line belonging to the 95th percentile values. This structure is particularly susceptible to range errors because it is located centrally within the anatomy and lies at the distal end of all five beam paths. When only seven range error points are considered, the corresponding dose values become disproportionately represented in the probability distribution. This leads to pronounced spikes and, consequently, an overestimation of the 95th-percentile dose values. When 9 range errors are considered instead of 7, the spikiness improves.

**Supplementary materials 4**

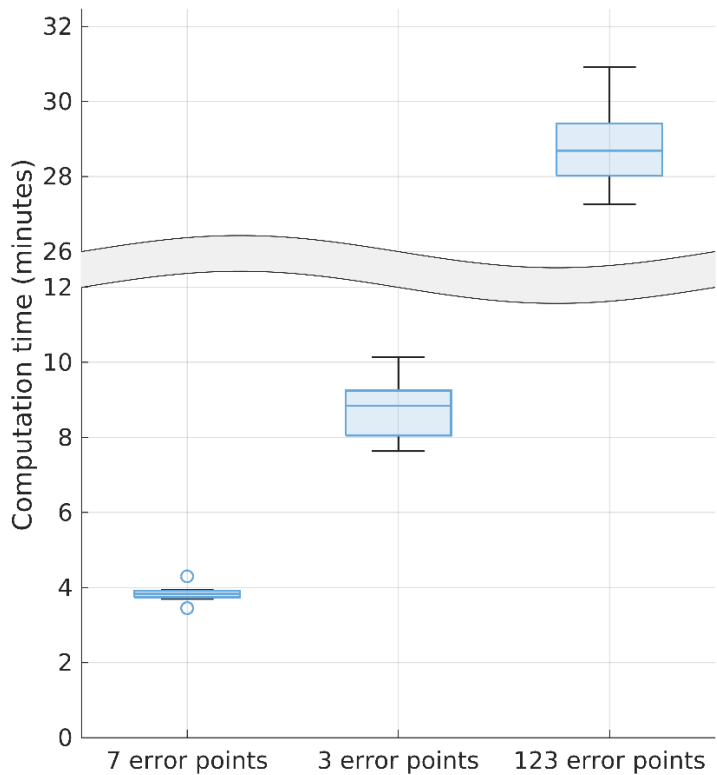

**Fig. 10.** Boxplots of the duration of probabilistic evaluation for 7, 33, and 123 included error scenarios ($n_{setup}$). Probabilistic evaluation encompasses the MC dose calculations and the creation of the probability distributions of the relevant DVH-parameters. Each boxplot contains 10 data points, corresponding to five patients for whom dose calculations were performed twice using $E_{max}=3\sigma$ and $E_{max}=4\sigma$. No systematic difference in computation time was observed between the two $E_{max}$ values.